\documentclass[5p,sort&compress,times,letterpaper]{elsarticle}
\usepackage{amsmath,amssymb}
\usepackage{upgreek}
\usepackage{graphicx}
\usepackage{hyperref}

\newcommand{\GF}{G_\text{F}}
\newcommand{\avg}[1]{\langle#1\rangle}
\newcommand{\Gaa}{G_\text{4b}}
\newcommand{\Gca}{G_\text{3a}}
\newcommand{\W}{\varOmega}

\newcommand{\rmd}{\mathrm{d}}
\newcommand{\rmi}{\mathrm{i}}

\newcommand{\pv}{\vec{p}}
\newcommand{\rv}{\vec{r}}
\newcommand{\vv}{\hat{v}}

\newcommand{\bfP}{\mathbf{P}}
\newcommand{\bfe}{\mathbf{e}}

\newcommand{\tr}{\text{tr}}
\newcommand{\diag}{\text{diag}}

\newcommand{\sfH}{\mathsf{H}}

\newcommand{\figscale}{0.6}

\begin{document}

\title{Dynamic fast flavor oscillation waves in dense neutrino gases}

\author{Joshua D.\ Martin}
\author{Changhao Yi}
\author{Huaiyu Duan}
\ead{duan@unm.edu}
\address{Department of Physics \& Astronomy, University of New
  Mexico, Albuquerque, New Mexico 87131, USA}

\begin{abstract}
  The flavor transformation in a dense neutrino gas can have a significant impact on the physical and chemical evolution of its surroundings. In this work we demonstrate that a dynamic, fast flavor oscillation wave can develop spontaneously in a one-dimensional (1D) neutrino gas when the angular distributions of the electron neutrino and antineutrino cross each other. Unlike the 2D stationary models which are plagued with small-scale flavor structures, the fast flavor oscillation waves remain coherent in the dynamic 1D model in both the position and momentum spaces of the neutrino. The electron lepton number is redistributed and transported in space as the flavor oscillation wave propagates, although the total lepton number remains constant. This result may have interesting implications in the neutrino emission in and the evolution of the compact objects such as core-collapse supernovae.

\end{abstract}



\maketitle

\section{Introduction}

Dense neutrino gases exist in the early universe and at the early epochs of some compact objects such as core-collapse supernovae and neutron star mergers. The transformation or oscillations between the electron-flavor (anti-)neutrinos and other neutrino species in these environments can have important consequences on their physical and chemical evolutions. However, the well-known flavor oscillation mechanisms through vacuum mixing or the Mikheev-Smirnov-Wolfenstein (MSW) effect
\cite{Wolfenstein:1977ue,Mikheev:1986if}
do not operate in such environments because of the presence of large matter densities. Through the neutrino-neutrino forward scattering
\cite{Fuller:1987aa,Notzold:1987ik,Pantaleone:1992xh}, the whole neutrino gas can experience flavor transformation collectively (e.g., \cite{Kostelecky:1993yt,Pastor:2001iu,Duan:2006jv};
see also \cite{Duan:2010bg,Chakraborty:2016yeg} for reviews). When the angular distributions of $\nu_e$ and $\bar\nu_e$ cross each other (and the other neutrino species have the same emission properties), the collective oscillations can occur on distance/time scales of $\sim \GF n_\nu$ \cite{Sawyer:2015dsa,Chakraborty:2016lct}, where $\GF$ and $n_\nu$ are the Fermi coupling constant and the neutrino density, respectively. The rapid progress in this research area may lead to a revolution in the current understanding of the supernova physics.

With a few exceptions \cite{Dasgupta:2017oko,Abbar:2018beu,Capozzi:2018clo}, the existing studies of the fast neutrino oscillations focus on the dispersion relations (DR) of the flavor oscillation waves and the associated instabilities which are derived from the linearized equations of motion (EoM) that govern the flavor transformation of the neutrino medium
\cite{Dasgupta:2016dbv,Izaguirre:2016gsx,Wu:2017qpc,Abbar:2017pkh,Capozzi:2017gqd,
Dasgupta:2018ulw,Airen:2018nvp,
Abbar:2018shq,Yi:2019hrp,Azari:2019jvr,Capozzi:2019lso}.
An important reason for the lack of research in fast neutrino oscillations in the nonlinear regime is the large dimensionality of the problem (which has a total of seven dimensions in the time, coordinate and momentum spaces). The only studies that investigate the nonlinear regime either make the assumption of the complete homogeneity \cite{Abbar:2018beu} or use a toy model with just a few (anti-)neutrino beams \cite{Capozzi:2018clo} or do both \cite{Dasgupta:2017oko}. In this work we study a one-dimensional (1D) neutrino gas of many neutrino momentum modes with the translation symmetries imposed along the $x$ and $y$ directions and the axial symmetry about the $z$ axis. Using this model we demonstrate how a small perturbation in an almost homogeneous flavor distribution can give rise to a dynamic, nonlinear flavor oscillation wave which redistributes and transports the electron lepton number across space.

\section{Equations of motion}
We consider the mixing between two neutrino flavors, $\nu_e$ and $\nu_\tau$ with $\nu_\tau$ being an appropriate linear combination of the physical $\nu_\mu$ and $\nu_\tau$. The flavor content of the neutrino medium at time $t$ and position $\rv$ can be described by the neutrino flavor density matrix $\rho_{\pv}(t,\rv)$ with $\pv$ being the momentum of the neutrino \cite{Sigl:1992fn}. The diagonal elements of $\rho$ (in the weak-interaction basis) give the number densities of the neutrinos in the corresponding flavors, and the off-diagonal elements are the coherences. The flavor density matrix $\bar\rho_{\pv}(t,\rv)$
of the antineutrino is defined similarly. In the absence of collisions, the neutrino flavor density matrix obeys the EoM (see, e.g., Ref.~\cite{Stirner:2018ojk} and the references therein)
\begin{align}
\rmi (\partial_t + \vv \cdot \vec{\nabla})
\rho_{\pv} = \left[
  \frac{\mathsf{M}^2}{2\varepsilon} + \sfH_\text{mat} +
  \sfH_{\nu \nu} ,
  \rho_{\pv}\right],
\label{eq:eom}
\end{align}
where $\varepsilon=|\pv|$, $\vv=\pv/\varepsilon$ and
$\mathsf{M}^2$ are the energy, velocity and the mass-square matrix of the
neutrino, respectively. In the above equation, $\sfH_\text{mat} =  \sqrt{2} G_\text{F} n_e \diag [1, 0]$ is the matter potential with $n_e$ being the net electron number density, and
\begin{align}
    \sfH_{\nu\nu} &= \sqrt{2} G_\text{F}
  \int\!(1-\vv\cdot\vv')(\rho_{\pv'} - \bar\rho_{\pv'})\,
  \frac{\rmd^3 p'}{(2\pi)^3}
\end{align}
is the neutrino potential.

We will focus on the fast oscillations for which the $\mathsf{M}^2$ in Eq.~\eqref{eq:eom} can be ignored apart from seeding the initial perturbations \cite{Sawyer:2015dsa, Chakraborty:2016lct}. We assume that neutrinos are in (almost) pure weak-interaction states with uniform densities initially and that $n_e$ is constant and uniform. For simplicity, we further assume that the neutrino gas possesses a perfect axial symmetry about the $z$ axis which is also a symmetry of the EoM. In this case, it is convenient to define the initial (angular) electron lepton number (ELN) distribution of the neutrinos to be \cite{Izaguirre:2016gsx}
\begin{align}
  G(u) = \frac{2\pi}{n_{\nu_e}}
  \int_0^\infty [(f_{\nu_e} - f_{\nu_\tau}) - (f_{\bar\nu_e} - f_{\bar\nu_\tau})]
  \frac{\varepsilon^2 \rmd\varepsilon}{(2\pi)^3},
\end{align}
where $u=v_z$ is the velocity component of the neutrino along the $z$ axis, and $f_\nu(\pv)$ and $n_\nu = \int f_\nu(\pv)\frac{\rmd^3 p}{(2\pi)^3}$ are the occupation number and the number density of the neutrino species $\nu$ initially. Because the dependence of the neutrino energy appears only through the vacuum mixing term in the EoM, the fast oscillations of both the neutrinos and antineutrinos of the same value of $u$ but different energies can all be represented by a single normalized flavor polarization vector $\bfP(t, z, u)$ which is related to the flavor density matrix by $P_i \propto \tr(\rho \sigma_i)$ ($i=1,2,3$) with $\sigma_i$ being the Pauli matrices. A polarization vector $\bfP=[0,0,1]$
indicates that the (anti)neutrino is in the electron flavor.
Such a flavor polarization vector has the EoM
\begin{align}
  (\partial_t + u \partial_z)\, \bfP =
  \left[\lambda\bfe_3 + \mu \int_{-1}^1 (1-u u')\, G' \bfP'\,\rmd u'\right]\times\bfP,
  \label{eq:eom-P}
\end{align}
where $\bfe_i$ ($i=1,2,3$) are the unit vectors in flavor space, $\lambda=\sqrt2 G_\text{F} n_{e}$ and $\mu = \sqrt2 G_\text{F} n_{\nu_e}$ are the strengths of the matter and neutrino potentials, respectively, and $G'$ and $\bfP'$ are the corresponding quantities that depend on $u'$.

The ELN distribution at the spacetime point $(t,z)$ is given by $G(u)P_3(t,z,u)$.
From Eq.~\eqref{eq:eom-P} one can easily deduce the ELN conservation law  \cite{Duan:2008fd}
\begin{align}
  \partial_t\avg{P_3} + \partial_z\avg{u P_3} = 0,
\end{align}
where $\avg{f}=\int_{-1}^1 G(u)f(u)\,\rmd u$ for an arbitrary function $f(u)$. Unlike in a homogeneous neutrino gas \cite{Hannestad:2006nj}, the ELN density $\avg{P_3}$ is no longer constant, and the ELN can be redistributed and transported in space.

In the rest of the paper, we measure the distance and time in the unit of $\mu^{-1}$ by setting $\mu=1$. We will also assume $\lambda=0$. For a finite and constant matter density, this is equivalent to choosing a reference frame which rotates about $\bfe_3$ with an angular frequency $\lambda$. Such a transformation does not change the value of $P_3$ which determines the conversion probability between the two neutrino flavors  \cite{Duan:2005cp}.

\section{Convective and absolute instabilities}

In the limit $P_\perp\ll1$, where $P_\perp$ is the magnitude of the component of $\bfP$ that is perpendicular to $\bfe_3$, Eq.~\eqref{eq:eom-P} can be linearized by dropping the terms of $\mathcal{O}(P_\perp^2)$ or higher. From the linearized EoM one can derive the DR $\W(K)$ of the collective oscillation wave \cite{Izaguirre:2016gsx} which is of the form
\begin{align}
  S(t, z, u) = P_1 -\rmi P_2 \propto e^{\rmi (K z - \W t)},
\end{align}
where $\W$ and $K$ are the frequency and wavenumber of the collective oscillation wave, respectively.
The neutrino gas possesses a flavor instability if there exists a DR branch with $\text{Im}(\W)>0$. In this case, $P_\perp$ grows exponentially with time and significant flavor transformation can ensue. The flavor instabilities can be either convective or absolute \cite{Capozzi:2017gqd}. For a convective instability, a perturbation moves away from its origin as it grows in both amplitude and extent. In contrast, the perturbation still embraces the point of the origin during the growth of an absolute instability \cite{Sturrock:1958zz}.

\begin{figure*}[ht!]
  \begin{center}
    $\begin{array}{@{}c@{\hspace{0.1in}}c@{}}
      \includegraphics*[scale=\figscale]{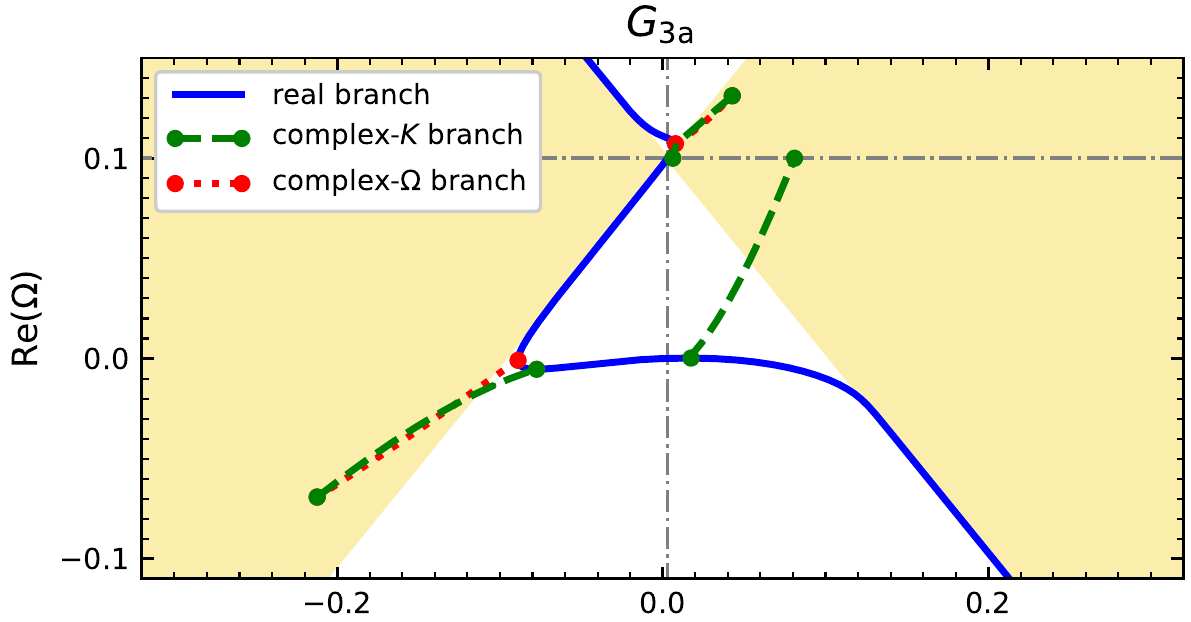} &
      \includegraphics*[scale=\figscale]{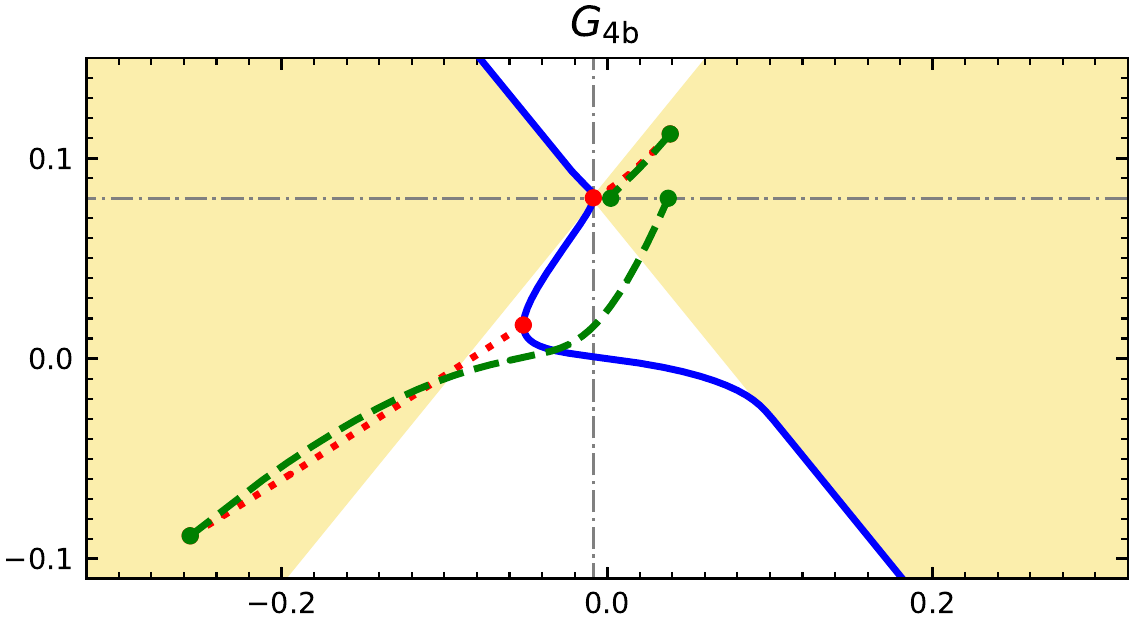} \\
      \includegraphics*[scale=\figscale]{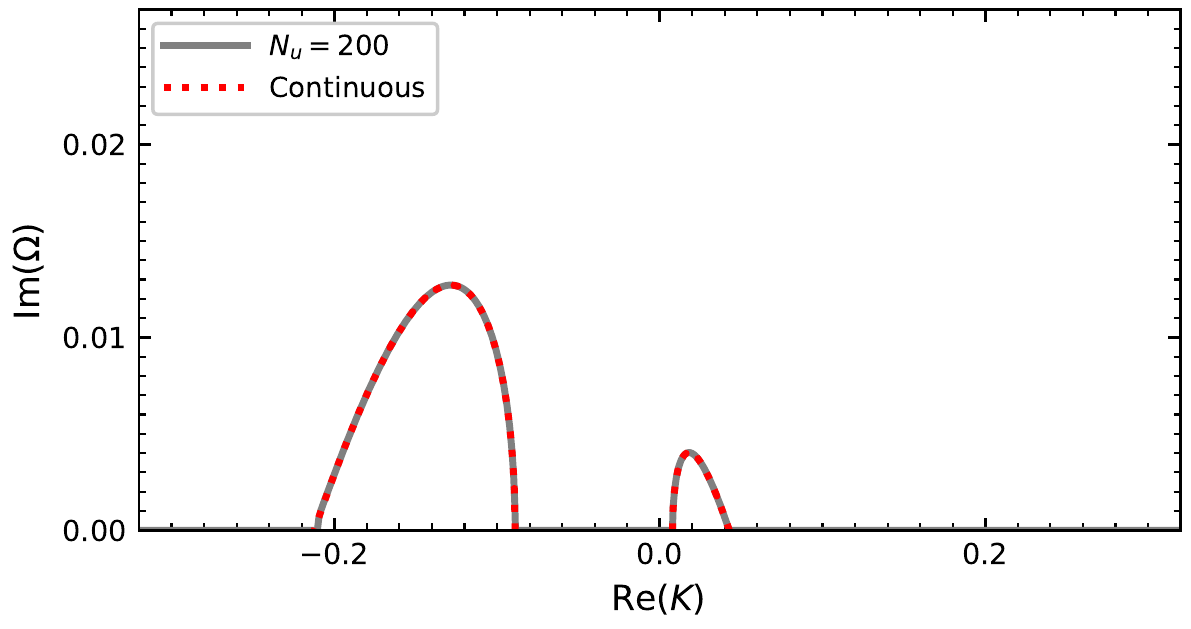} &
      \includegraphics*[scale=\figscale]{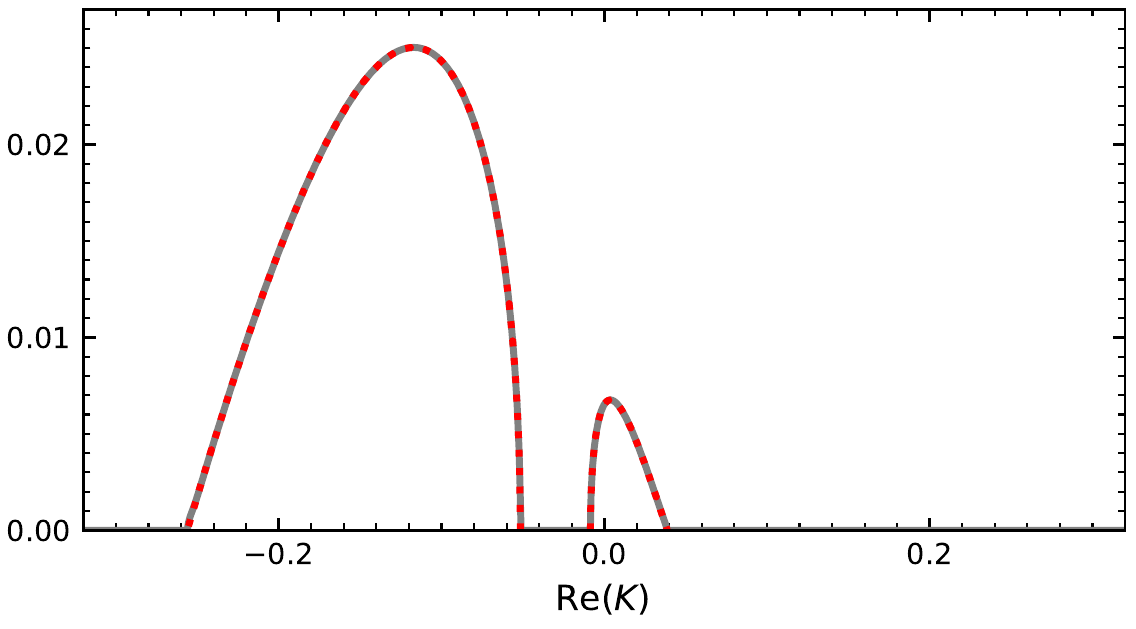}
      \end{array}$
  \end{center}
  \caption{The various DR branches of the fast flavor oscillation waves in the dense neutrino gases with the initial ELN distributions $\Gca$ (left) and $\Gaa$ (right), respectively. The results are obtained using the continuous ELN distributions except for the solid curves in the bottom panels which are calculated with 200 discrete angle bins. Unlike in the early references such as \cite{Izaguirre:2016gsx,Yi:2019hrp}, the values of $\W$ and $K$ shown here are not shifted by $\int_{-1}^1 G(u)\rmd u$ and $\int_{-1}^1 u G(u)\rmd u$ (which are shown as the horizontal and vertical dot-dashed lines, respectively). The so-called forbidden regions which are avoided by the real DR branches are shown as the shaded regions.}
  \label{fig:DR}
\end{figure*}

To demonstrate these two kinds of instabilities, we consider two ELN distributions, $\Gca$ and $\Gaa$, both of which are of the form
\begin{align}
G(u) = g(u, 0.6) - \alpha g(u, 0.53),
\end{align}
where $g(u, \xi) \propto \exp[-(u-1)^2/2\xi^2]$ with normalization $\int_{-1}^1 g\,\rmd u=1$, and $\alpha=0.9$ for $\Gca$ and $0.92$ for $\Gaa$, respectively. (These distributions are shown as the solid curves in Fig.~\ref{fig:conversion}.) We computed the (collective) DR branches for both distributions and show them in Fig.~\ref{fig:DR}.

The ELN distributions of the dynamic model studied in this work can be classified into six categories according to the analytic properties of their DRs with no ELN crossing for the distributions in category I and deepest crossings in category VI \cite{Yi:2019hrp}. Distributions $\Gca$ and $\Gaa$ fall into categories III and IV, respectively, and illustrate the difference between the convective and absolute instabilities.
Distribution $\Gca$ has a convective instability which is associated with a pair of complex-$\W$ and complex-$K$ DR branches (shown as the dotted and dashed curves, respectively) in the lower left quadrant (delineated by the  horizontal and vertical dot-dashed lines) of the upper left panel of Fig.~\ref{fig:DR}. These two branches originate from the same (critical) point inside the forbidden region (shown as the shaded region) and end at two separate (critical) points on the same real DR branch (shown as the solid curve). A separate complex-$K$ branch connects from the real branch to the horizontal dot-dashed line from below. As the ELN crossing becomes deeper and $\Gca$ becomes $\Gaa$, these two complex-$K$ branches fuse together and bypass the real branch (on a different Riemann plane). This indicates that the convective instability has become absolute. Similar process has occurred for the complex-$K$ branches above the horizontal dot-dashed line as the ELN distribution evolve from category II to III which produce similar absolute instabilities for both $\Gca$ and $\Gaa$. (See Ref.~\cite{Yi:2019hrp} for more details.)

To demonstrate the natures of the instabilities, we solved Eq.~\eqref{eq:eom-P} with a variant of the numerical code which solves the stationary 2D neutrino line model \cite{Martin:2019kgi}. The calculations were performed on a 1D lattice of $24,000$ points equally spaced on the $z$ axis in a periodic box of size $L=1200$. We assumed the initial condition
\begin{align}
\bfP(0,z,u)=[\epsilon(z), 0, \sqrt{1-\epsilon^2(z)}]
\label{eq:P-init}
\end{align}
with
\begin{align}
  \epsilon(z) = \epsilon_0\, e^{-(z-z_0)^2/50},
  \label{eq:eps}
\end{align}
where $\epsilon_0=10^{-6}$ and $10^{-7}$ for the cases with $\Gca$ and $\Gaa$, respectively, and $z_0=L/2$. The width of the initial perturbation is chosen in such a way that it is well resolved on the lattice in the coordinate space and that it encompasses both complex-$\W$ branches in the Fourier space. We used 200 equally spaced angle bins for $u\in[-1,1]$ which reproduce the correct $\text{Im}(\W)$ in the linearized flavor stability analysis (lower panels of Fig.~\ref{fig:DR}). We verified that the Fourier transform of $\bfP(t,z,u)$ has the same exponential growth rate as what is predicted by this analysis. Unlike stationary 2D models
\cite{Chakraborty:2015tfa, Abbar:2015mca}, however, there are no unstable spurious modes in this model.

\begin{figure*}[ht!]
  \begin{center}
    $\begin{array}{@{}c@{\hspace{0.1in}}c@{}}
      \includegraphics*[scale=\figscale]{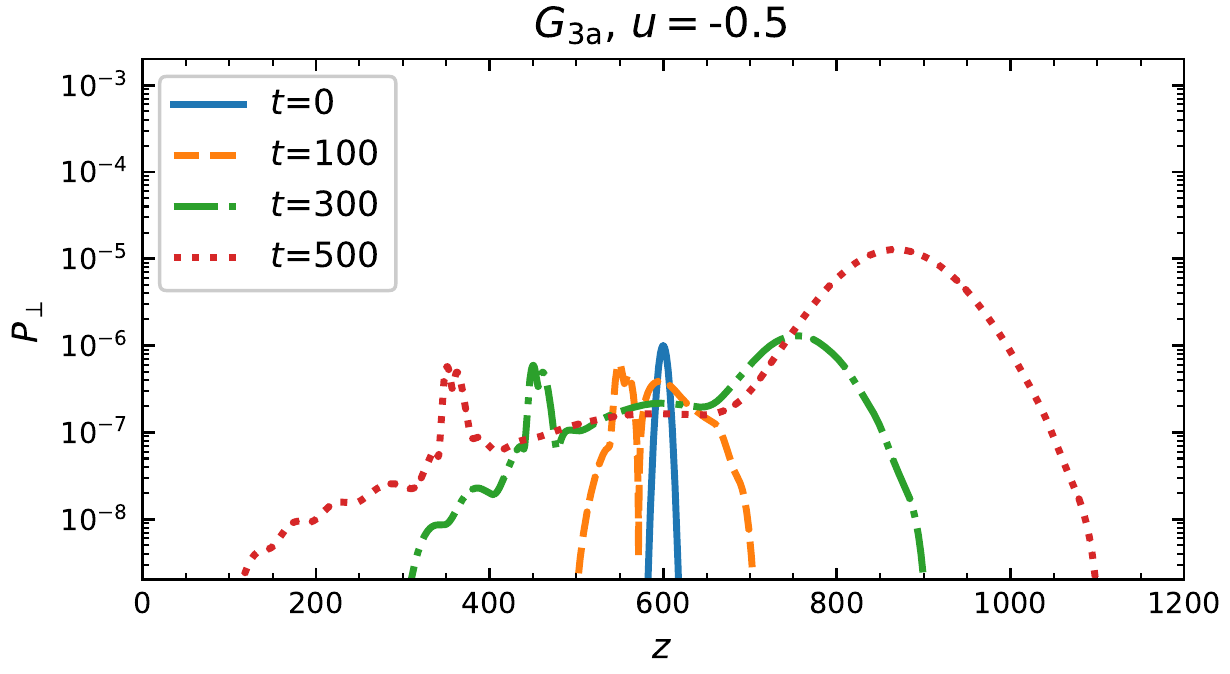} &
      \includegraphics*[scale=\figscale]{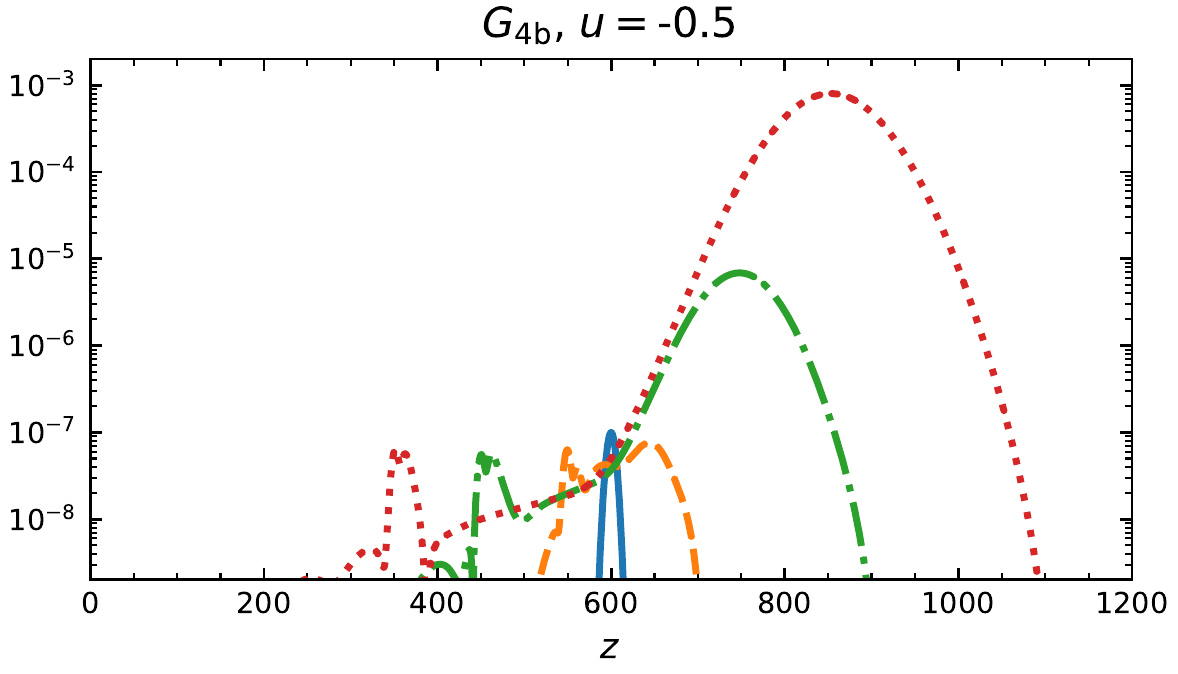}
      \end{array}$
  \end{center}
  \caption{The perpendicular components $P_\perp$ of the neutrino polarization vectors of $u=-0.5$ as functions of the position $z$ at various times $t$ (as labeled). The neutrino gases are confined within a periodic box of size $L=1200$ with similar initial perturbations as specified in Eqs.~\eqref{eq:P-init} and \eqref{eq:eps} but with different magnitudes: $\epsilon_0=10^{-6}$ for the gas with the $\Gca$ distribution (left panel) and $10^{-7}$ for $\Gaa$ (right panel).}
  \label{fig:inst}
\end{figure*}

In Fig.~\ref{fig:inst} we show $P_\perp(t, z, -0.5)$ for a few snapshots in both cases. For the gas with $\Gca$, the initial perturbation splits into two at later times. One part of the perturbation is advected toward the left with an almost constant peak amplitude and a velocity $V\approx -1$. The other part of the perturbation moves toward the right while growing in both amplitude and extent. The leftward advection of one part of the perturbation is due to the section of the real DR branch with the group velocity $V=\rmd\W/\rmd K\approx -1$. The growth and expanding of the other part of the perturbation is because of the convective instability which was previously explained. The absolute instability of $\Gca$ is dwarfed by the convective instability and cannot be clearly identified in this calculation. The behavior of the neutrino gas with $\Gaa$ is similar to that with $\Gca$ with an important difference. Because the growing perturbation is caused by an absolute instability in this case, it expands both right and (slightly) left, although its peak still moves  rightward. (This is clearly seen by comparing, e.g., the green dot-dashed curves and the red dotted ones in the figure.)

\section{Nonlinear regime}

\begin{figure*}[ht!]
  \begin{center}
    $\begin{array}{@{}c@{\hspace{0.1in}}c@{}}
      \includegraphics*[scale=\figscale]{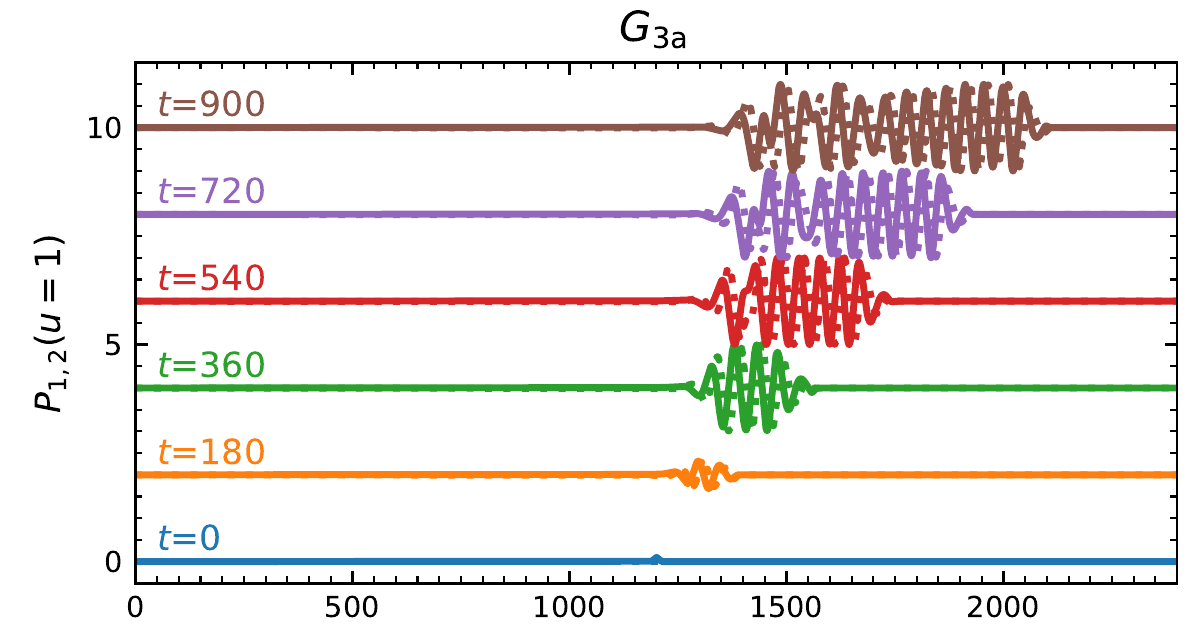} &
      \includegraphics*[scale=\figscale]{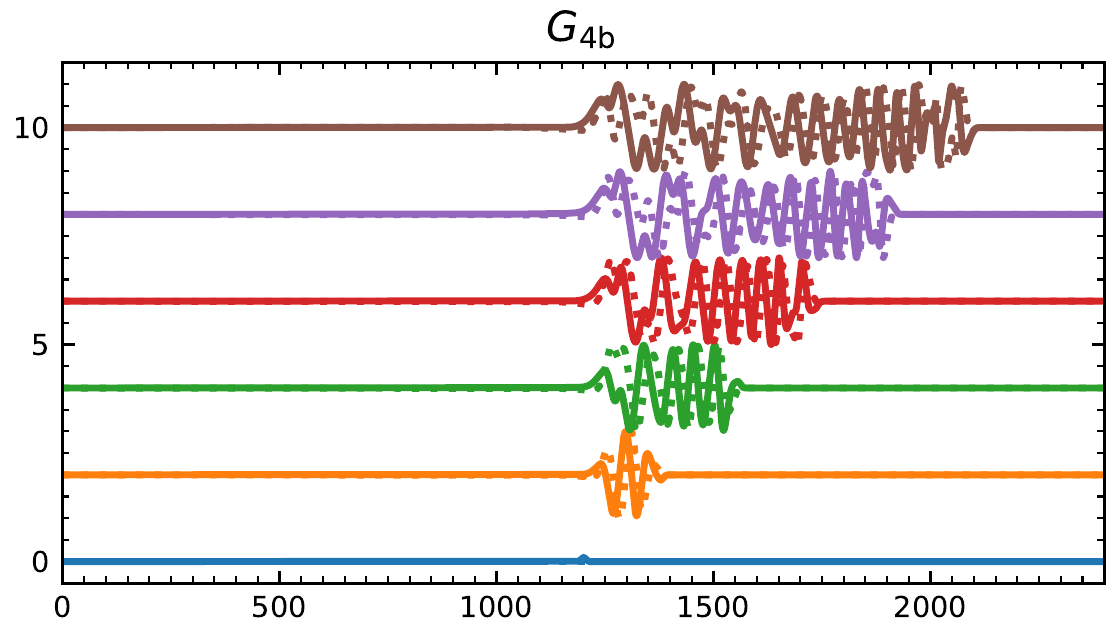} \\
      \includegraphics*[scale=\figscale]{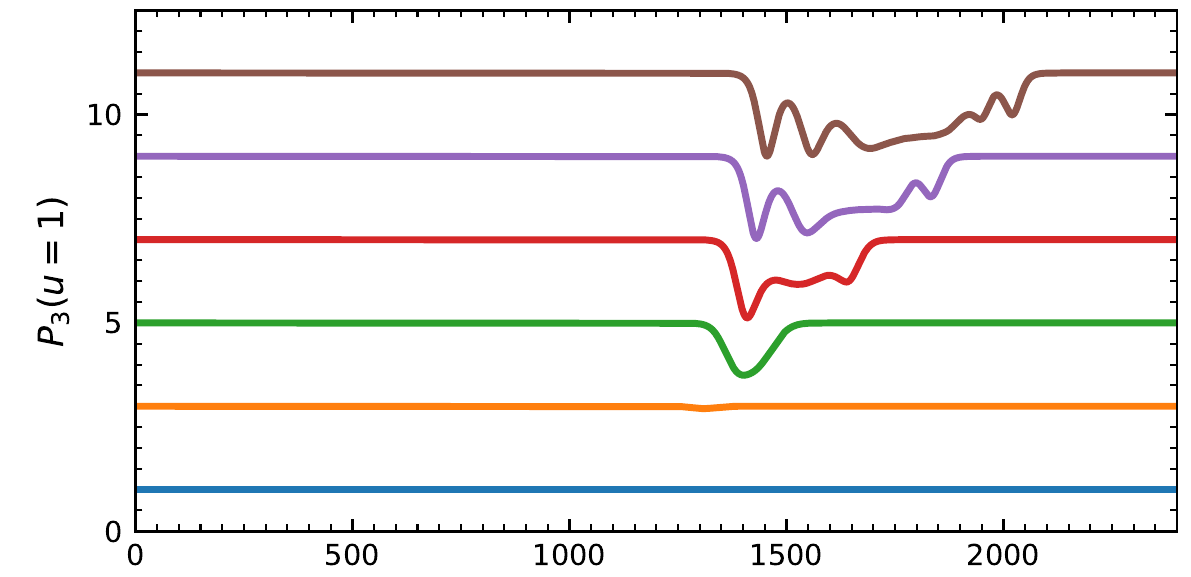} &
      \includegraphics*[scale=\figscale]{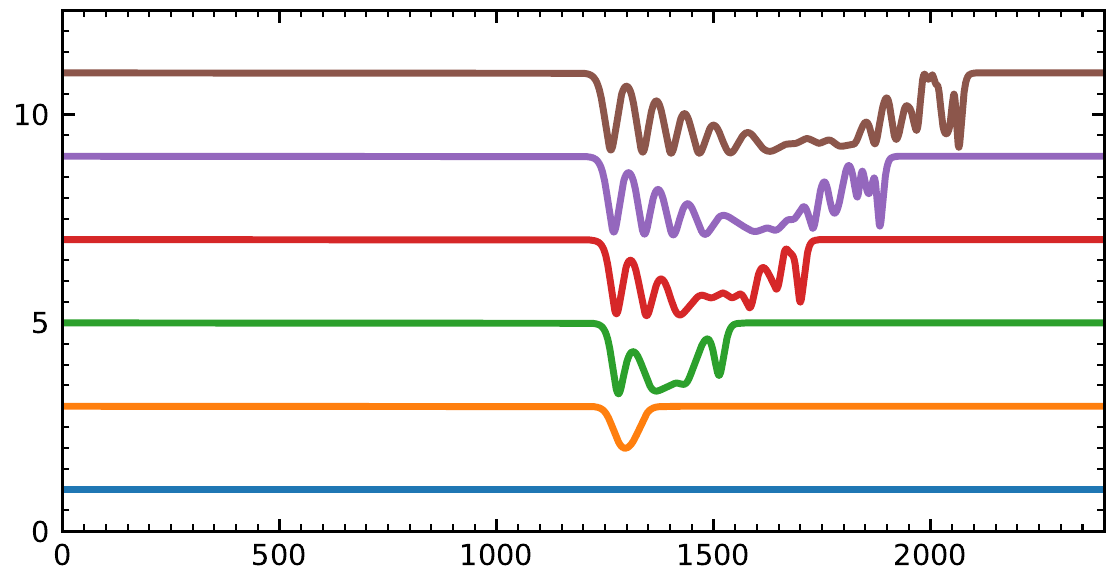} \\
      \includegraphics*[scale=\figscale]{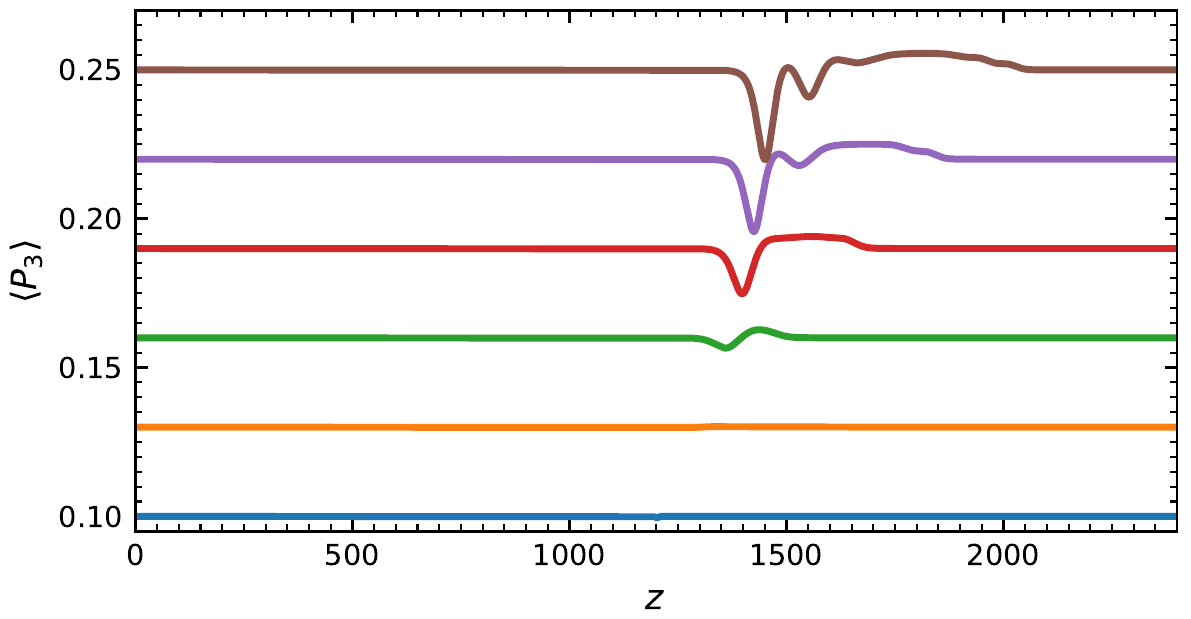} &
      \includegraphics*[scale=\figscale]{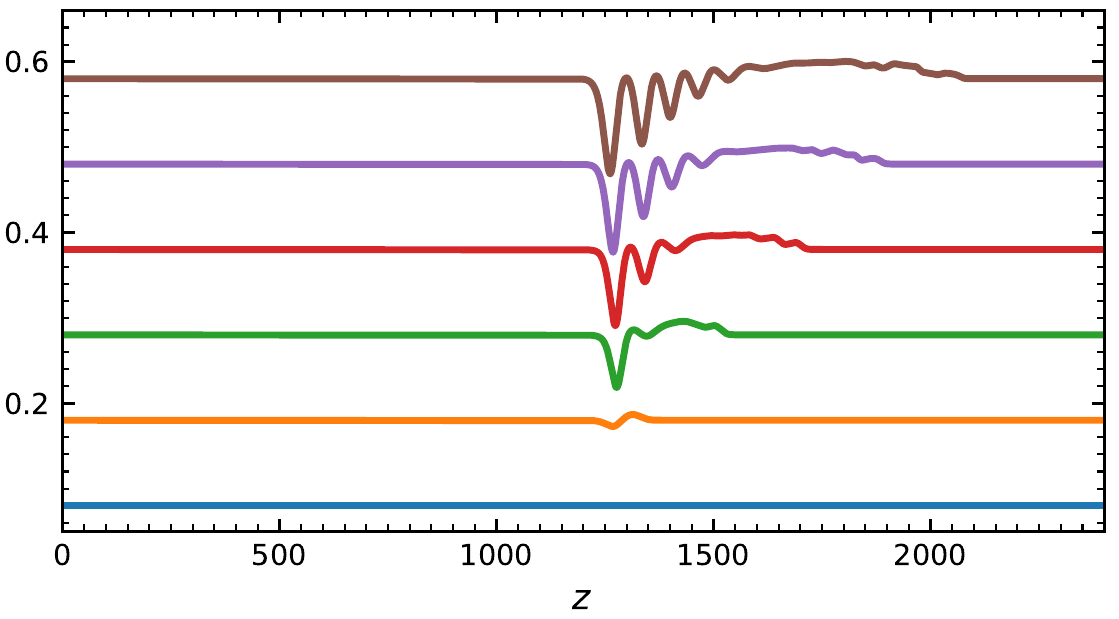}
      \end{array}$
  \end{center}
  \caption{The polarization vectors ($P_1$ as solid curves and $P_2$ as dotted curves in the top panels, and $P_3$ in the middle panels) of the neutrino gases for the right going beams and the ELN density $\avg{P_3}$ (bottom panels) as functions of position $z$ at various times $t$ (as labeled). The curves are offset from each other (by 2 units in the top and middle panels, 0.03 in the bottom left panel, and 0.1 in the bottom right panel) for clarity. The setups of the calculations are similar to those in Fig.~\ref{fig:inst} except with a larger box size $L=2400$ and a larger initial perturbation ($\epsilon_0=0.1$ for both $\Gca$ and $\Gaa$).}
  \label{fig:waves}
\end{figure*}

\begin{figure*}[ht!]
  \begin{center}
    $\begin{array}{@{}c@{\hspace{0.1in}}c@{}}
      \includegraphics*[scale=\figscale]{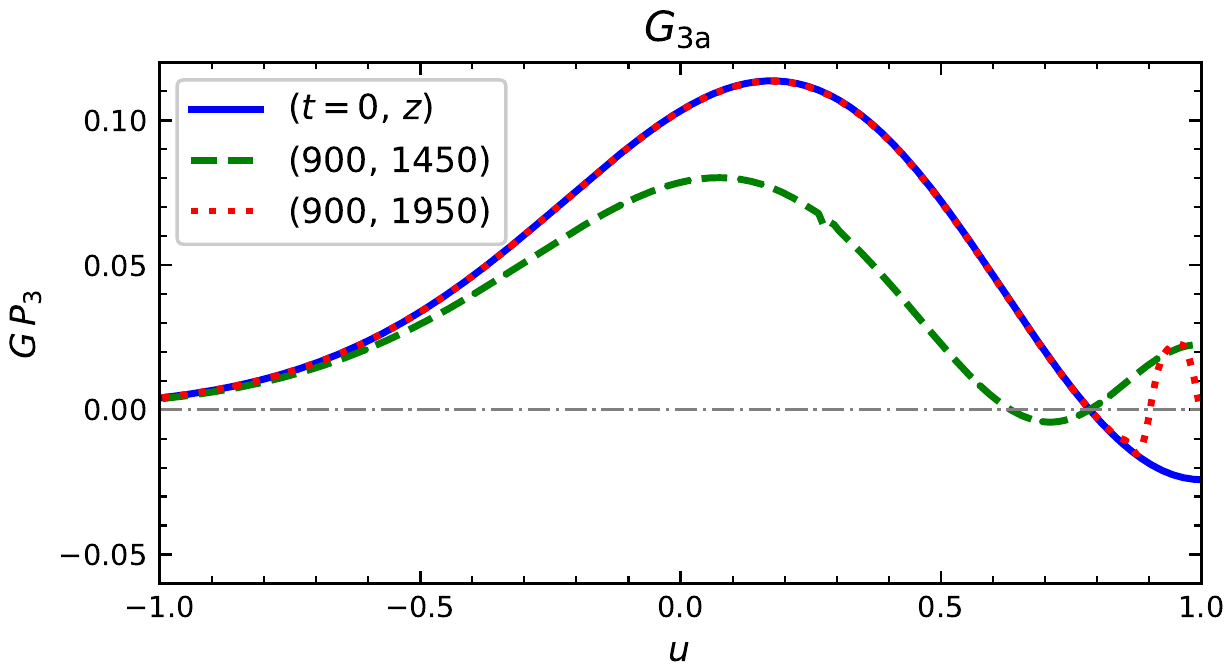} &
      \includegraphics*[scale=\figscale]{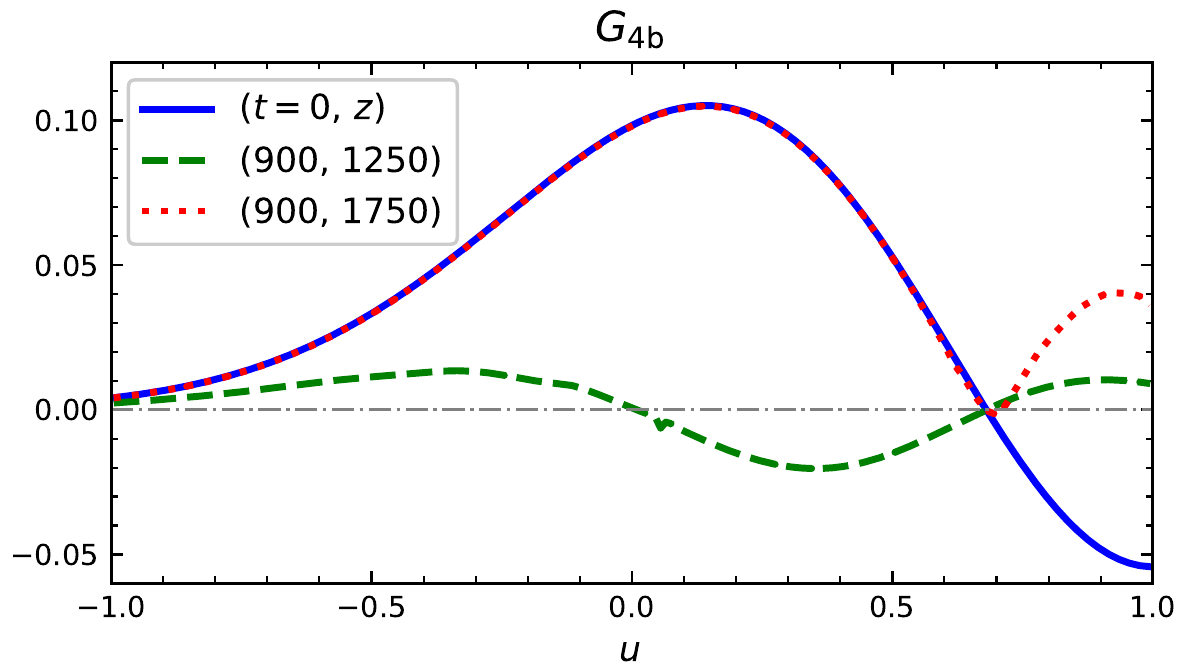}
      \end{array}$
  \end{center}
  \caption{The comparison of the initial ELN distributions ($t=0$, solid curves) and those at a couple of different spacetime points $(t,z)$ (dashed and solid curves) for the calculations described in Fig.~\ref{fig:waves}.}
  \label{fig:conversion}
\end{figure*}

\begin{figure*}[ht!]
  \begin{center}
    $\begin{array}{@{}c@{\hspace{0.1in}}c@{}}
      \includegraphics*[scale=\figscale]{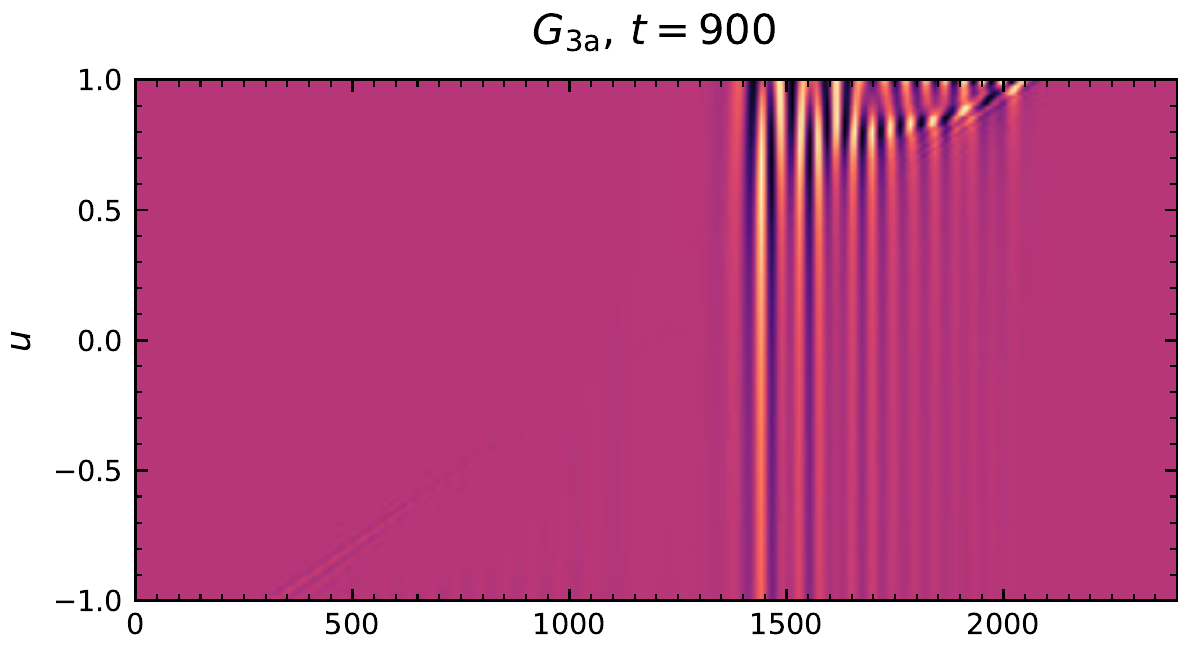} &
      \includegraphics*[scale=\figscale]{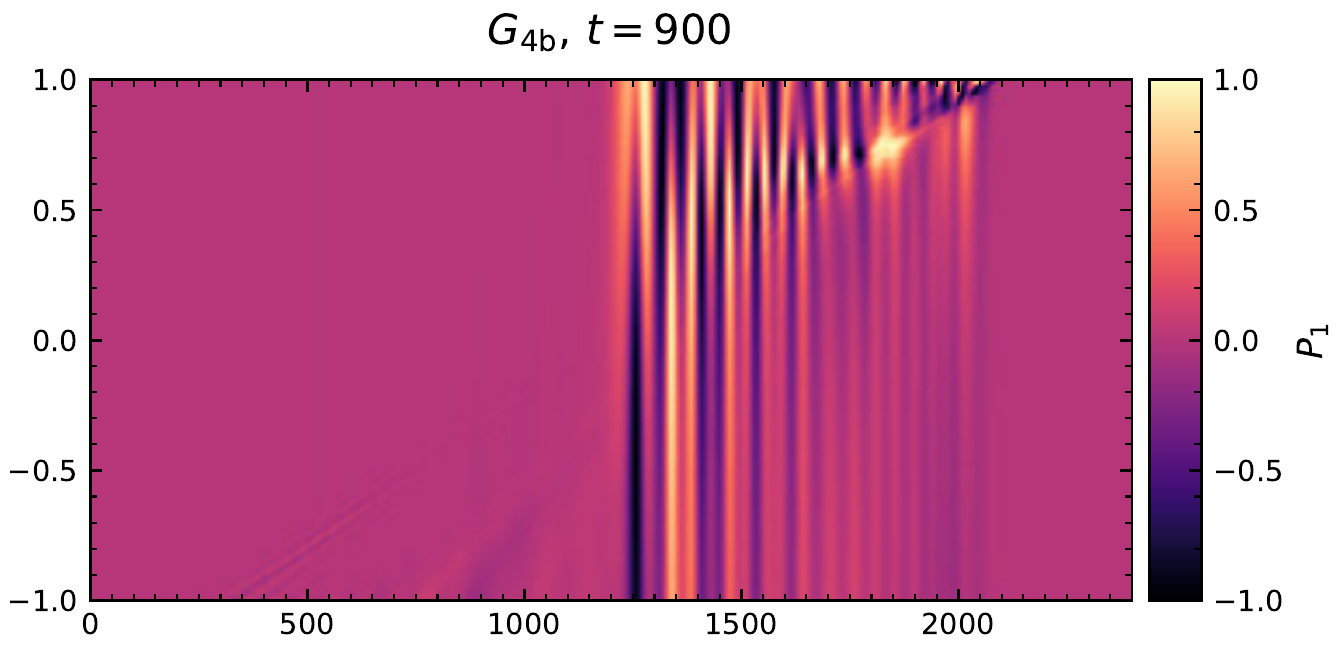} \\
      \includegraphics*[scale=\figscale]{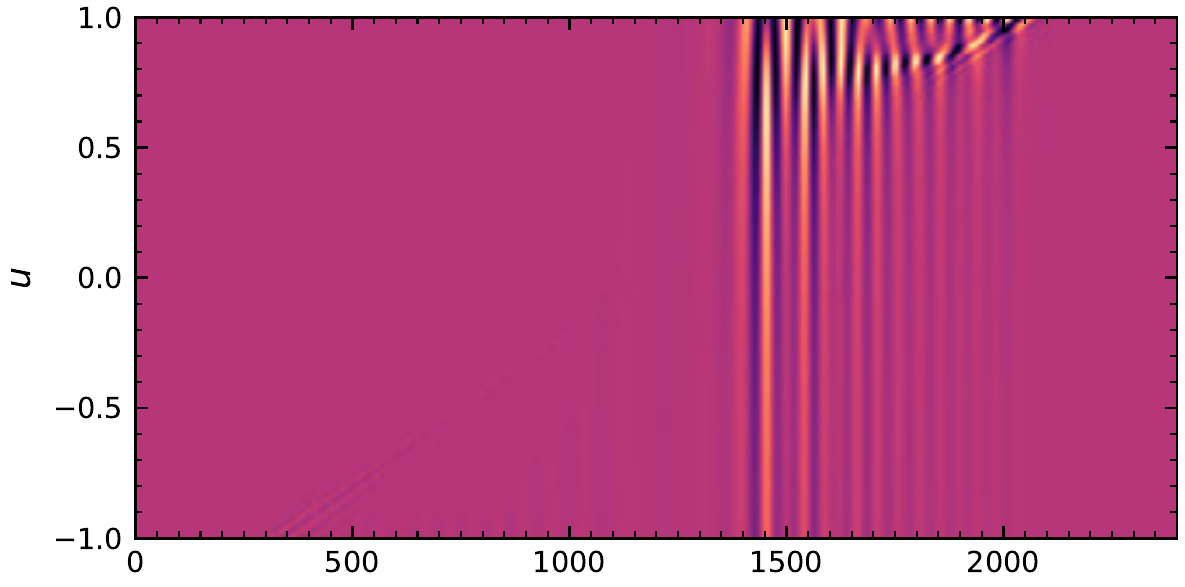} &
      \includegraphics*[scale=\figscale]{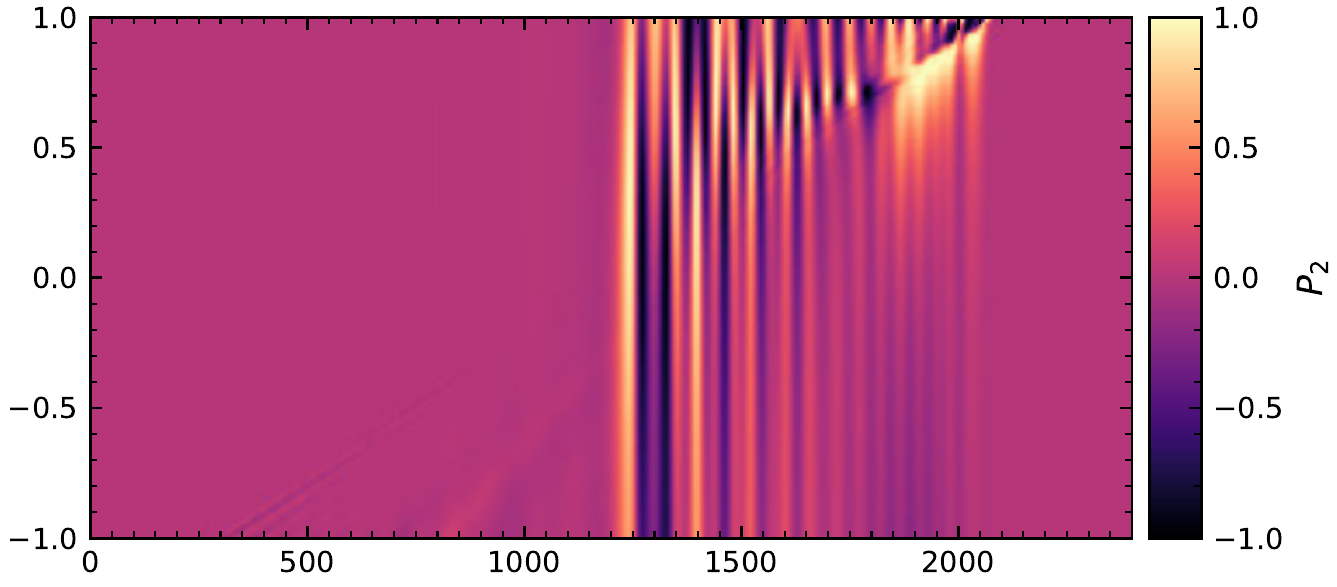} \\
      \includegraphics*[scale=\figscale]{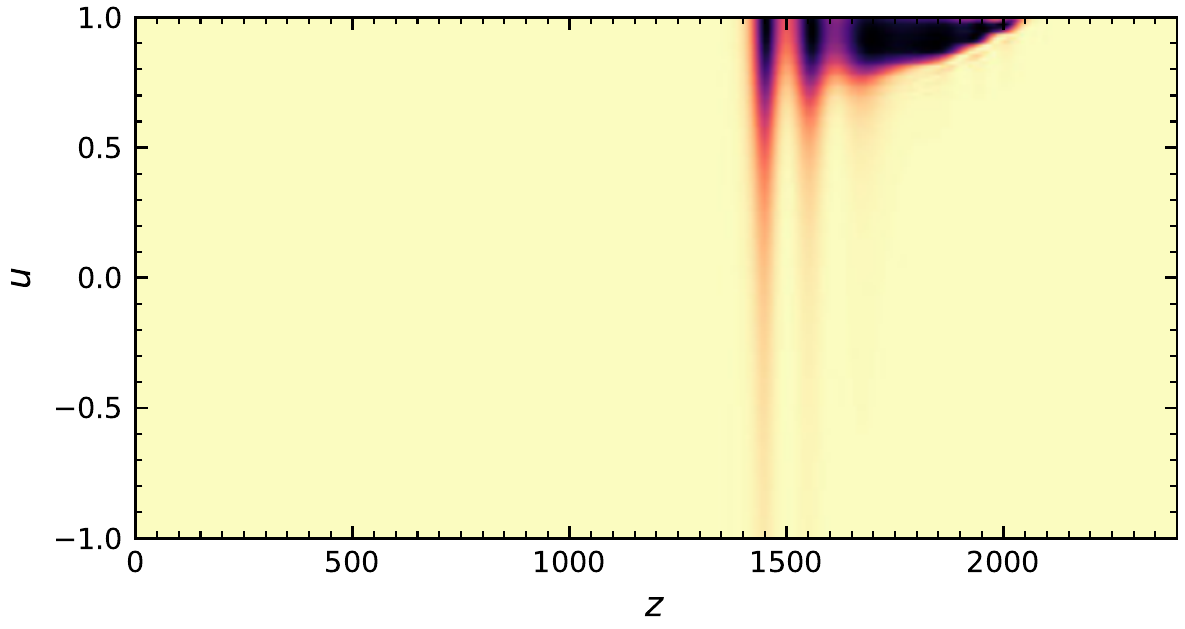} &
      \includegraphics*[scale=\figscale]{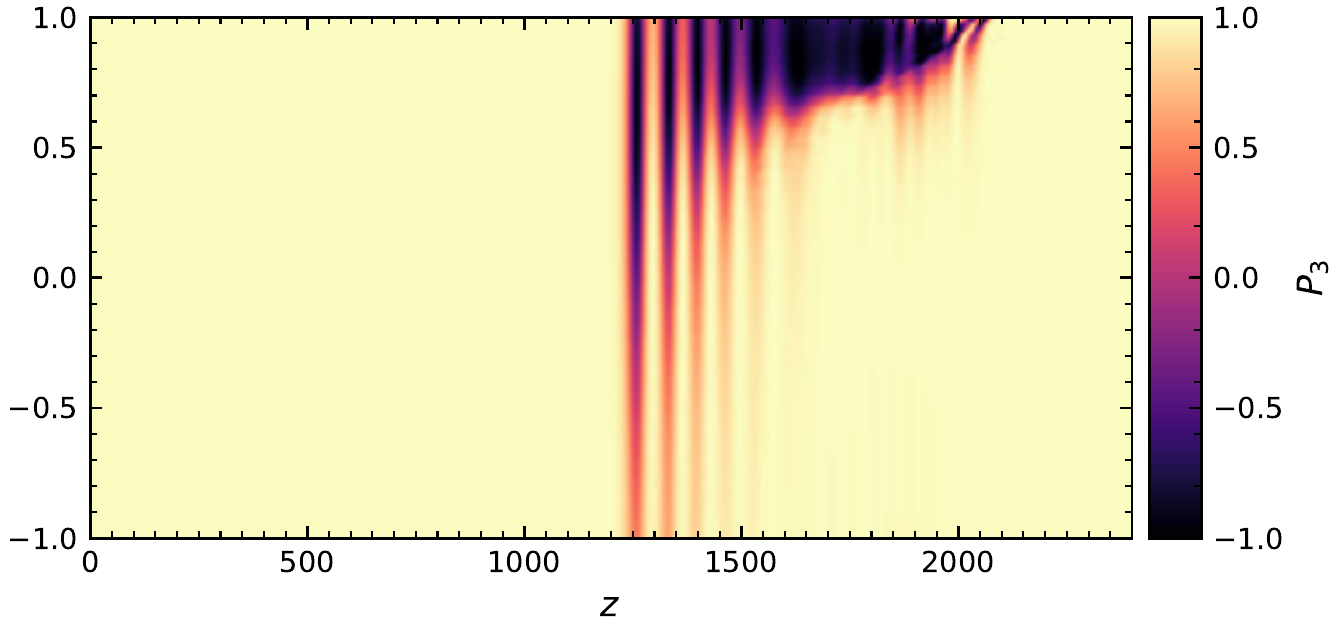}
      \end{array}$
  \end{center}
  \caption{The three polarization components, $P_1$ (top panels), $P_2$ (middle panels), and $P_3$ (bottom panels), of the neutrino gases in the calculations described in Fig.~\ref{fig:waves} at $t=900$.}
  \label{fig:snapshot}
\end{figure*}

To study the fast oscillations in the nonlinear regime, we again consider the neutrino gases with the initial ELN distributions $\Gca$ and $\Gaa$. We assume the same initial conditions described by Eqs.~\eqref{eq:P-init} and \eqref{eq:eps} but with $\epsilon_0=0.1$ for both cases. We also use a larger box with $L=2400$ resolved by $48,000$ lattice points.

In the top and middle panels of Fig.~\ref{fig:waves} we show a few snapshots of each components of the polarization vectors $\bfP$ of the right going neutrino beams. In both cases, the initial small perturbations grow into  flavor oscillation waves which are described by the precession of $\bfP$ in the plane perpendicular to the flavor axis $\bfe_3$. The regions in which the waves exist extend with time as they propagate.
Similar to the linear regime, the wave region moves slowly away from the origin of the perturbation in the gas with the $\Gca$ distribution, but continues to embrace the origin of the perturbation in the gas with the $\Gaa$ distirbution.

In the bottom panels of Fig.~\ref{fig:waves} we show the ELN density $\avg{P_3}$ for the same snapshots as in the top and middle panels. Instead of having a uniform spatial distribution, the lepton number is transported and redistributed as the neutrino oscillation waves propagate. Specifically, $\avg{P_3}$ develops a dip near the end of the wave, an extensive plateau in the forepart of the wave, and some oscillations in between. In Fig.~\ref{fig:conversion} we show the ELN distribution $G P_3$ at $t=900$ for two representative positions in each case, one in the dip, and the other on the plateau. On the plateau the flavor conversion occurs almost completely in the forward direction where $G(u)<0$, while in the dip the flavor conversion occurs across the entire angular distribution.

In Fig.~\ref{fig:snapshot} we show the components of $\bfP$ for all neutrino beams at $t=900$. Unlike in the stationary 2D models which are plagued with small-scale flavor structures \cite{Mirizzi:2015fva,Mirizzi:2015hwa,Martin:2019kgi}, a coherent flavor oscillation pattern is clearly seen in both neutrino gases.

\section{Discussion and conclusions}

In this work we studied the fast flavor oscillation waves in two 1D neutrino gases with the ELN distributions that are similar to what have been found in the neutrino decoupling region of some supernova models \cite{Abbar:2018shq}.
By solving the EoM numerically that governs the fast oscillations, we  demonstrated that the initial ELN distributions which satisfy the corresponding criteria given in Ref.~\cite{Yi:2019hrp} are indeed associated with convective and absolute instabilities in the linear regime. Because of these instabilities, a small perturbation in an initial flavor distribution that is almost uniform in space can develop into a fast oscillation wave which both propagate and extend in space.
Although the distinction between the convective and absolute instabilities is relative and depends on the choice of the reference frame \cite{Briggs:1964}, it is of practical value for supernova physics because one almost always chooses the proto-neutron star to be at rest.

Although the model that we used in this work has the same number of total dimensions as the stationary 2D models \cite{Mirizzi:2015fva,Mirizzi:2015hwa,Martin:2019kgi}, their behaviors are entirely different in the nonlinear regime. While the stationary 2D models are plagued with spurious instabilities and teemed with small-scale flavor structures, there is no spurious instability in the dynamic 1D model, and the flavor oscillations remain coherent for the neutrinos with different momenta and at different locations.

Our findings can have interesting implications for supernova physics. The neutrino emission properties and even the explosion dynamics of the core-collapse supernova can be affected as fast oscillations induce large-amplitude flavor conversions and transport the lepton numbers. Following the previous studies of the fast neutrino oscillations in the linear regime \cite{Dasgupta:2016dbv,Tamborra:2017ubu,Abbar:2018shq}, this work calls for more self-consistent supernova simulations such as \cite{Nagakura:2017mnp} with detailed information of the angular distributions of the neutrinos emitted. In the future one should also consider the feedback to the electron distribution due the fast neutrino flavor conversions.

\section*{Acknowledgments}
We thank S.~Abbar for the useful discussion.
We acknowledge the support by the US DOE
NP grant No.\ DE-SC0017803 at UNM.

\bibliographystyle{elsarticle-num}
\bibliography{fsw}

\end{document}